\documentclass[12pt]{article}

\usepackage{scicite}
\usepackage{graphicx}
\usepackage{braket}
\usepackage{amsmath}
\usepackage{color}
\usepackage{times}

\usepackage{graphicx}           
\usepackage{epstopdf}
\usepackage{subfigure}
\DeclareGraphicsExtensions{.eps,.ps}
\usepackage{pstricks}           
\usepackage{dcolumn}            
\usepackage{bm}                 
\usepackage{url}

\newenvironment{sciabstract}{%
\begin{quote} \bf}
{\end{quote}}

\newcounter{lastnote}

\usepackage{braket}
\usepackage{bbm}

\title{\vspace{-3cm}Hitherto unrecognized intermolecular Coulombic decay mechanism in gases}
\author
{Alan G. Falkowski$^{1,2}$, Alexander I. Kuleff$^{1}$, Lorenz S. Cederbaum$^{1,\ast}$\\
 \\
\normalsize{$^{1}$ Theoretische Chemie, Physikalisch-Chemisches Institut (PCI), Universit\"at Heidelberg,}\\
\normalsize{69120 Heidelberg, Germany}\\
\normalsize{$^{2}$ Instituto de Física “Gleb Wataghin”, Universidade Estadual de Campinas,}\\ 
\normalsize{13083-859 Campinas, São Paulo, Brazil}\\
\normalsize{$^\ast$ \textbf{Corresponding author. E-mail: Lorenz.Cederbaum@pci.uni-heidelberg.de}}\\
\\
}

\date{}

\begin{document}

\baselineskip24pt

\maketitle

\begin{sciabstract}
 Excited atoms and molecules can utilize their excess energy to ionize a neighboring system by a process named interatomic and intermolecular Coulombic decay (ICD). ICD is ultrafast, in the femtosecond regime, and has many modes of appearance. Ample applications of ICD have been reported spanning a wide range of fields and it is expected to be ubiquitous in nature. Essentially all the investigations on ICD were for weakly bound systems, like clusters and fluids. We demonstrate that, unexpectedly, ICD can be efficiently active in atomic and molecular gases in spite of the very large distances between the units. We uncover the underlying mechanism, which differs from that prevailing in weakly bound systems. The dynamics of ICD in gases is analyzed. The results considerably broaden the impact of ICD and open the gateway to new kinds of applications.   
\end{sciabstract}


\section{Introduction} \label{sec:Introduction}

Interatomic and intermolecular Coulombic decay (ICD) \cite{Giant,ICD_Review_CR} is by now a theoretically and experimentally well investigated ultrafast relaxation process of excited neutral or charged electronic systems.  The excited system utilizes its excess energy to ionize a neighboring system provided that the excess energy is sufficiently high to ionize the neighbor. Because the final state lies in the continuum, energy conservation is always fulfilled and the process is resonant. Consequently, both the initially excited system and the neighbor can be atoms or molecules, and the corresponding timescale of ICD is in the femtosecond regime even in the presence of a single neighbor \cite{ICD_Review_CR}. Importantly, ICD becomes faster the more neighbors are present and this has been theoretically predicted and experimentally observed \cite{Giant,ICD_more_N_2,ICD_Review_CR,ICD_second_layer_Fasshauer_2016,ICD_more_N_2023}. 

There are ample applications of ICD spanning a wide range of fields varying from quantum halo systems with an extreme mean separation between the atoms~\cite{Nico_He_Dimer,Exp_He_Dimer,ICD_LiHe_Anael}, quantum fluids~\cite{ICD_quantum_liquids_review} to quantum dots and wells~\cite{ICD_QW_Nimrod,ICD_QD_Annika,ICD_QD_Nimrod}. Of general interest is that ICD is of potential relevance in radiation damage and for molecules of biological interest~ \cite{21st_Centuary,ICD_Till_Water,ICD_Uwe_Water,ICD_liquid_water,ICD_Bio_Stoychev,Uwe_Bio_Review,ICD_RA_Nature,ICD_RA_Nature_Exp,Dorn_alpha_particles,Stumpf16a,Radiation_Damage_ETMD,Radiation_Damage_ETMD_2,Heavy_Ion_ICD_Shenyue}.

Until recently, all the investigations on ICD were for weakly bound systems, where hydrogen bonds and van der Waals interactions prevail, like in clusters and fluids. Very recently, experiments on unbound molecules have been reported where the ions detected have been attributed to ICD~\cite{ICD_in_unbound_pyridines}. A laser pulse of low intensity excites a gas of pyridine molecules with a photon energy where three photons are needed to ionize a single molecule. Nevertheless, ions of pyridine monomers and even of dimers and trimers have been detected. Their appearance has been attributed to collective ICD, where energy is collectively transferred to ionize a unit once three excited molecules are in relatively close proximity. This can lead to the growth of molecules which enhances the importance of the process in various contexts~\cite{ICD_in_unbound_pyridines,ICD_in_unbound_PANH}.

The reported experiments on unbound molecules have motivated us to pose the more general question: can ICD be active in a low-density gas of atoms and molecules where the distances between these units are large? In this work we shall demonstrate that the answer is positive and ICD can be rather efficient, and, in addition, we shall uncover the underlying mechanism. This substantially broadens the scope of ICD and is expected to open the door to  new kinds of applications.

\section{Results and Discussion} \label{sec:Results&Discussion}

\subsection{The basic scenario}\label{sec:The basic scenario}

\begin{figure}[!h] 
	\centering
	\includegraphics[width=0.5\textwidth]{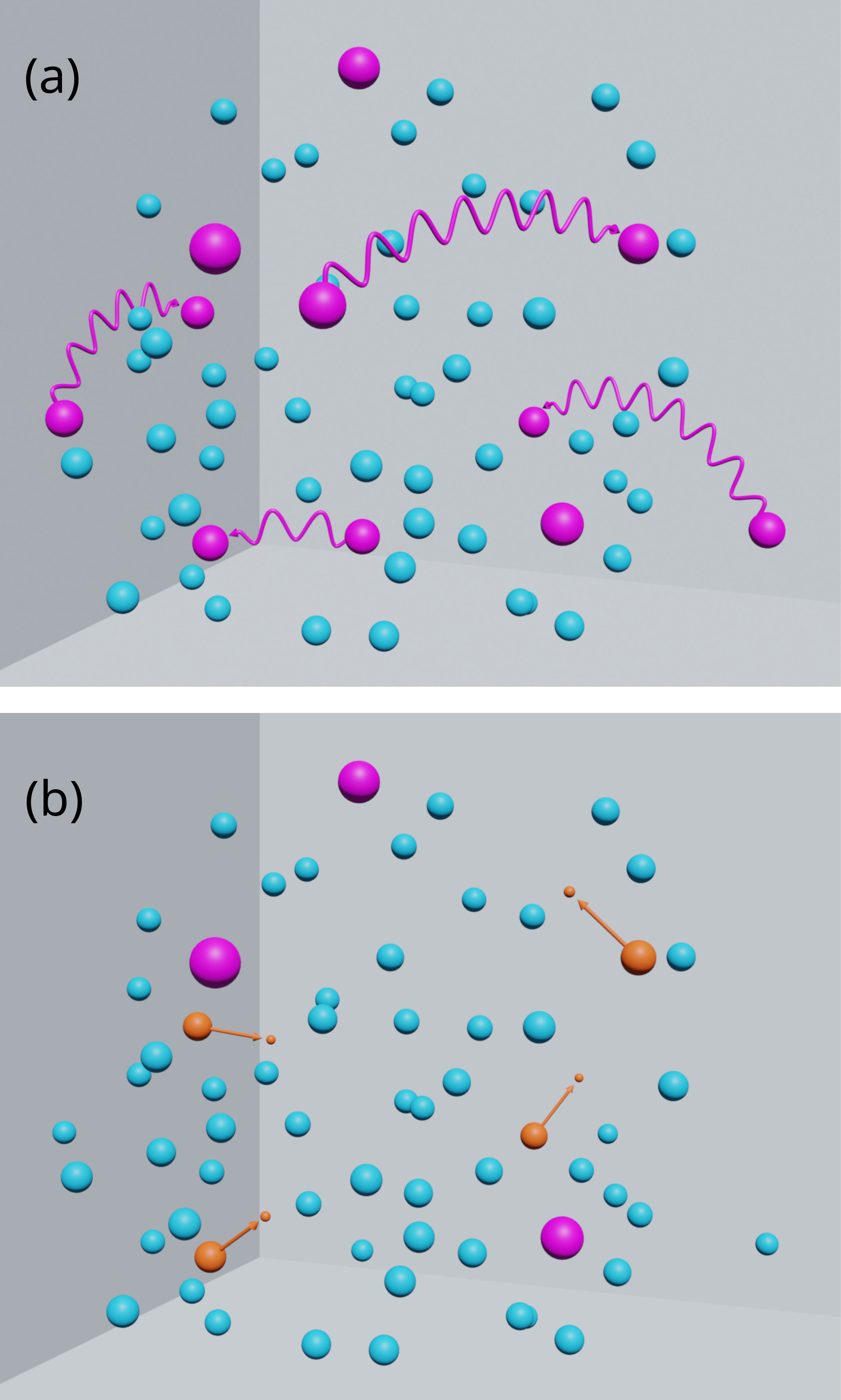}
	\caption{Schematic picture of ICD in gases. Panel A) Neutral species in their electronic ground state (small cyan balls) and electronically excited species (larger magenta balls). Each excited species can potentially transfer its excess energy to ionize another excited species. In other words, any excited species can be an acceptor as well as a donor. An energy transfer is by a so called virtual photon~\cite{Averbukh04} and indicated by a wavy line. Four possible transfers are explicitly shown which might take place at an arbitrary small time interval. Panel B) After the transfers indicated in panel A have taken place, the four donors have turned into ground state species and the four acceptors into ions (small orange balls). The respective four ICD electrons are also indicated.}
	\label{Schematic_Cartoon_ICD_Gas}
\end{figure}

We consider a gas of atoms or molecules $A$ of which at time $t=0$ a number $N^*(0)$ are electronically excited. In the absence of ICD, this number will decay exponentially as a function of time $t$ via the radiative rate $\gamma_{rad}$. Our goal is to compute the impact of ICD. If active at all, the ICD pathway would be that of multiply excited species predicted in \cite{PhysRevLett.105.043004} and measured in clusters~\cite{ICD_Multiply_Excited_Exp_1,ICD_Multiply_Excited_Exp_2,ICD_Multiply_Excited_Exp_3}. In this pathway, one of the excited species  $A^*$ transfers its excess energy and ionizes another excited species: $A^* + A^* \rightarrow A + A^+$. Figure~\ref{Schematic_Cartoon_ICD_Gas} visualizes schematically the situation in the gas. Initially, there are many neutral and less excited species, and potentially any excited one can play the role of a donor and transfer its energy to ionize another one which is then the acceptor. Then, in time, the number of neutral species $N(t)$ in their ground state increases accordingly and ions produced by ICD appear.

ICD in weakly bound systems is triggered by the Coulomb interaction between the electrons of the donor and those of the acceptor. The interaction is expanded in inverse powers of the distance $R$ between the donor and acceptor, and the leading term of the ICD rate takes on the appearance~\cite{PhysRevLett.105.043004}
\begin{align}\label{ICD_rate_Coulomb}
	\gamma_{_{ICD}} = \frac{3}{4\pi}\left(\frac{c}{\omega}\right)^4\frac{\gamma_{rad}\sigma^*_{pi}}{R^6},
\end{align}
where $c$ is the speed of light, $\hbar\omega$ is the transferred energy and $\sigma^*_{pi}$ is the photoionization cross section of the acceptor. The superscript $^*$ should remind one that the acceptor is $A^*$, i.e., an electronically excited atom or molecule. Notice that as an excited species can be a donor as well as an acceptor, the total ICD rate of a pair of excited species is $2\gamma_{_{ICD}}$. 

In weakly bound systems, the donor-acceptor distances typically amount to several \r{A}, while in gases we rather have $\mu$m as a relevant scale. We show below that the ICD rate is rather negligible in gases unless accidentally two excited species come close to each other. Consequently, one has to go beyond Coulomb interaction. Due to the finite speed of light, retardation effects appear which enlarge the distance range in which ICD can take place. The resulting ICD rate has been predicted by employing QED within the dipole approximation~\cite{Retardation_ICD_1} and, more recently, by relativistic theory employing the Breit interaction~\cite{Relativistic_Energy_Transfer}. If the transferred energies are large, the relativistic theory predicts additional relevant effects~\cite{Relativistic_Energy_Transfer,ICD_Attosecond_Transfer}. For small transferred energies both approaches provide identical results giving
\begin{align}\label{ICD_rate_Retardation}
	\gamma_{_{ICD}} = \frac{\gamma_{rad}\sigma^*_{pi}}{4\pi}\left[3 \left(\frac{c}{\omega}\right)^4\frac{1}{R^6} + \left(\frac{c}{\omega}\right)^2\frac{1}{R^4} + \frac{1}{R^2}\right].
\end{align}
For weakly bound systems, the contribution of the additional terms to the ICD rate is very interesting, but generally small. We stress that the above expressions are valid at large distances. At intermolecular distances as
in weakly bound systems like clusters and liquids, the true power of ICD lies in finding that its rate can be orders of magnitude larger than that
predicted by the above formulas~\cite{Averbukh04}.

To be specific, we choose gas conditions similar to those reported in an explicit experiment~\cite{ICD_in_unbound_pyridines}. The gas volume is 0.125~cm$^3$ which is the volume of the interaction region reported and we vary the pressure around the reported one of 0.032~Pa (1~at = 10$^5$~Pa). At room temperature, 0.032~Pa implies 10$^{12}$ atoms or molecules and we choose 10\% of them to be excited as reported in the experiment.

Owing to the large number of particles present in the gas, we employ rate equations to compute the time evolution of the numbers $N^*(t)$ of excited species, $N(t)$ of species in the ground state and $N^+(t)$ of ions for different initial numbers $N^*(0)$ of excited ones. The rate equations are derived in the section Methods and are collected in Eq.~(\ref{ICD_rate_equations}). As described above, the excited species decay by radiation emitting a photon and by ICD. In each ICD event, two excited species are converted into a ground state one and into an ion, thereby emitting an ICD electron. Each excited species can contribute to the total ICD rate via the pairwise rate as in Eq.~(\ref{ICD_rate_Retardation}) with all the other excited ones. However, due to the presence of other species, one has to take attenuation into account which weakens the rate exponentially with the distance between the communicating excited species of a pair~\cite{Retardation_ICD_1,Relativistic_Energy_Transfer}, see Methods. To avoid averaging over many realizations of a gas, we have distributed the excited species isotropically in the volume. In this way, the shortest distance between two excited species is initially the average distance and grows further in due time as the number of excited species decreases. Consequently, there are no fluctuations in the gas which can bring excited species close to each other, and we may consider our computed ion yield to be a lower bound. 

What species to choose? As can be seen in the expression (\ref{ICD_rate_Retardation}) for the rate $\gamma_{_{ICD}}$, the photoionization cross section $\sigma^*_{pi}$ of the acceptor in its excited state is needed. The knowledge of this quantity is rather scarce. There are calculations available for the rare gas atoms Ne and Ar~\cite{popova2022spectroscopic,ojha1983photoionisation} and for the CO molecule~\cite{ruberti2014total}. We have thus chosen these species in order to present a proof of principle for ICD in gases. Expression~(\ref{ICD_rate_Retardation}) exhibits the radiative rate $\gamma_{rad}$ of a donor which makes clear that we should preferentially choose a singlet excited state. For Kr and Xe $\sigma^*_{pi}$ have been determined for triplets and we would like to complement our study by including these atoms too using these quantities as estimates for the respective singlets. All the data required for the calculations for these species are collected in Tables S1 and S2 in the Supplementary Information (SI). 

\begin{figure}[!h] 
	\centering
	\includegraphics[width=1.0\textwidth]{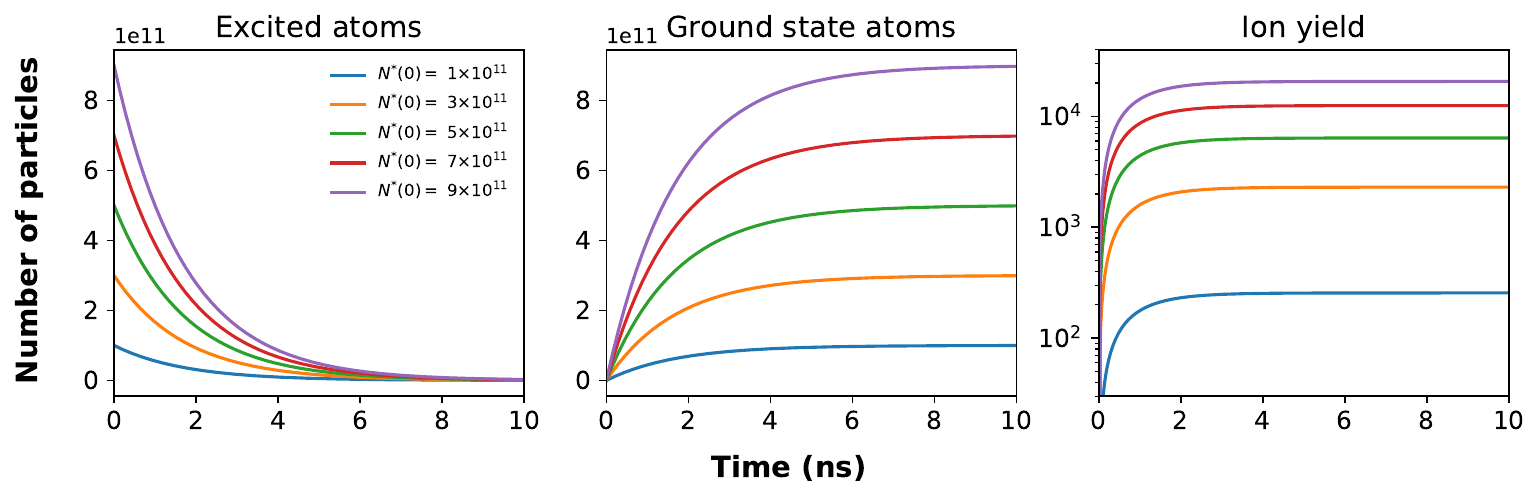}
	\caption{A Ne gas in a volume of 0.125 cm$^3$ with 10\% of excited atoms evolves in time. Shown are the results for different numbers of atoms corresponding to a pressure of 0.032 to 0.288~Pa at room temperature. The excited atoms can decay by radiation and by ICD. In each ICD event one excited atom is transferred to its ground state and one is ionized. The left panel depicts the evolution of the number $N^*(t)$ of excited Ne atoms, the middle panel shows the number $N(t)$ of Ne atoms in their ground state created by radiative decay and ICD and the right panel reports the ion yield $N^+(t)$ produced by ICD.}
	\label{Number_Particles_vs_Time}
\end{figure}

Let us start with Ne which is the least favorable for ICD of all the above mentioned examples. The excited state is characterized by [He]2s$^2$2p$^6 \rightarrow$  [He]2s$^2$2p$_{1/2}^53s^1$. The time evolution of the gas is shown in Figure~\ref{Number_Particles_vs_Time} for several numbers of initially excited Ne atoms. It is seen in the left panel that the number $N^*(t)$ of excited states decreases the faster the larger $N^*(0)$ is. This observation is discussed below. The number of ground state atoms grows in time owing to the radiative decay of the excited ones as well as due to ICD. Since both pathways are open only while excited atoms are present, this number saturates as $N^*(t)$ vanishes, see middle panel. Particularly interesting is the evolution of the ion yield shown in the right panel. The production of Ne ions is only possible by ICD and finding hundreds and even thousands of ions is a proof that ICD is operative in gases. 

To better understand the involved time scales, we realize that ICD becomes weaker as time proceeds because the number of excited species decreases by radiative decay as well as by ICD itself and, in addition, the distances between the remaining excited ones typically grow. Every excited species can undergo ICD with the other $N^*(t)-1$ ones and this defines its ICD rate and by summing over all excited species one obtains the total ICD rate $\Gamma_{_{ICD}}$ of the gas (see Methods for details). As usual, $\hbar/\Gamma_{_{ICD}}$ defines the time scale of the ICD decay, which we here call ICD time.

\begin{figure}[!h] 
	\centering
	\includegraphics[width=1.0\textwidth]{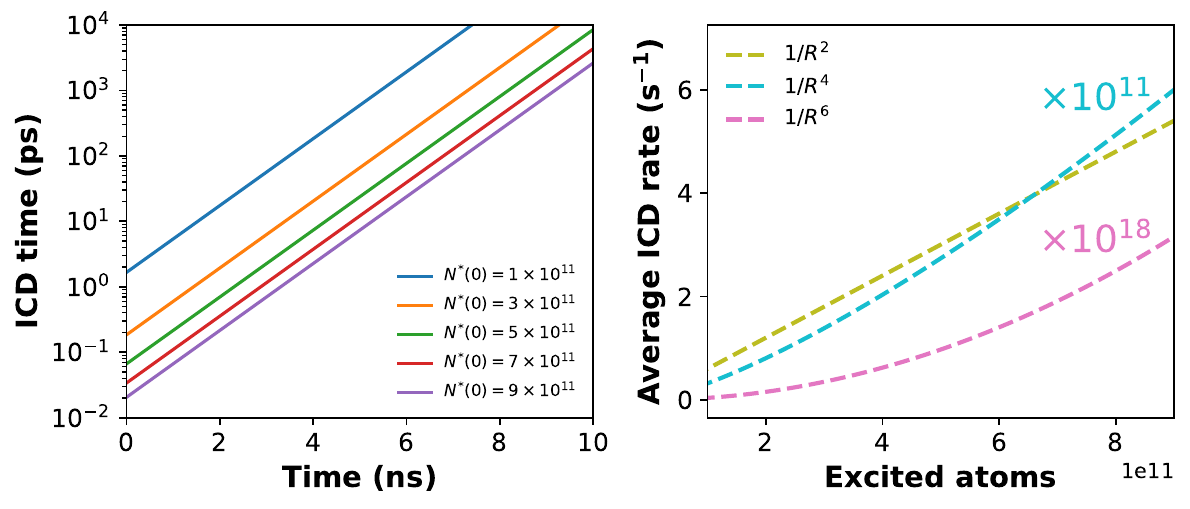}
	\caption{The ICD time and the average ICD rate for the Ne gases as in Figure~\ref{Number_Particles_vs_Time}. The total ICD rate of a gas $\Gamma_{_{ICD}}$ is determined by summing up all possible ICD rates between the excited species of the gas taking into account attenuation due to the gas (see Methods for details). As usual, $\hbar/\Gamma_{_{ICD}}$ defines the time scale of the ICD, which is here named ICD time. The total ICD rate and the ICD time are temporal quantities because the number of excited atoms decreases in time. The left panel shows the ICD time on a logarithmic scale as function of time. It is seen that the ICD time increases exponentially as time proceeds. ICD is initially very fast (ps and sub ps) and dominates the decay by radiation (1.7~ns radiative lifetime of the excited Ne atom). The situation reverses at later times, but by then the ion yield has already saturated (see right panel of Figure~\ref{Number_Particles_vs_Time}). The average ICD rate, i.e., $\Gamma_{_{ICD}}/N^*$ at $t=0$, is depicted in the right panel for the different Ne gases and broken into its three contributions. It is seen that all terms except of retardation can be neglected.}
	\label{ICD_Time_and_Rate}
\end{figure}

We have computed the total ICD rate and the ICD time for the Ne gas. Note that as the number $N^*(t)$ of excited atoms decreases in time, these are temporal quantities. The ICD time is shown in the left panel of Figure~\ref{ICD_Time_and_Rate} as a function of time for the same initial numbers $N^*(0)$ as in Figure~\ref{Number_Particles_vs_Time}. Obviously, the ICD is extremely fast initially: 2~ps for $N^*(0)=10^{11}$ and 20~fs for $N^*(0)=9 \times 10^{11}$. Clearly, the ICD becomes slower as time proceeds and arrives at the ns regime after about 6-10~ns. The radiative lifetime of an excited Ne is 1.7~ns, and we can now understand the fast rise of the ion yield seen in Figure~\ref{Number_Particles_vs_Time}. ICD is much faster at short times than radiative decay which dominates the overall decay at later times.      

Having found that ICD is rather efficient in Ne gases, we are now in the position to analyze the impact of the Coulomb interaction between the electrons of the donors and those of the acceptors in the gas. For that purpose we have computed the average ICD rate $\Gamma_{_{ICD}}/N^*$ for the various Ne gases investigated and broke the values down into their contributions due to Coulomb and due to retardation. The results depicted in the right panel of Figure~\ref{ICD_Time_and_Rate} dramatically demonstrate that only retardation is responsible for ICD in the gases. Remember that it is the Coulomb interaction which is responsible for ICD in weakly bound systems and retardation is very minor there. 

\begin{figure}[!h] 
	\centering
	\includegraphics[width=1.0\textwidth]{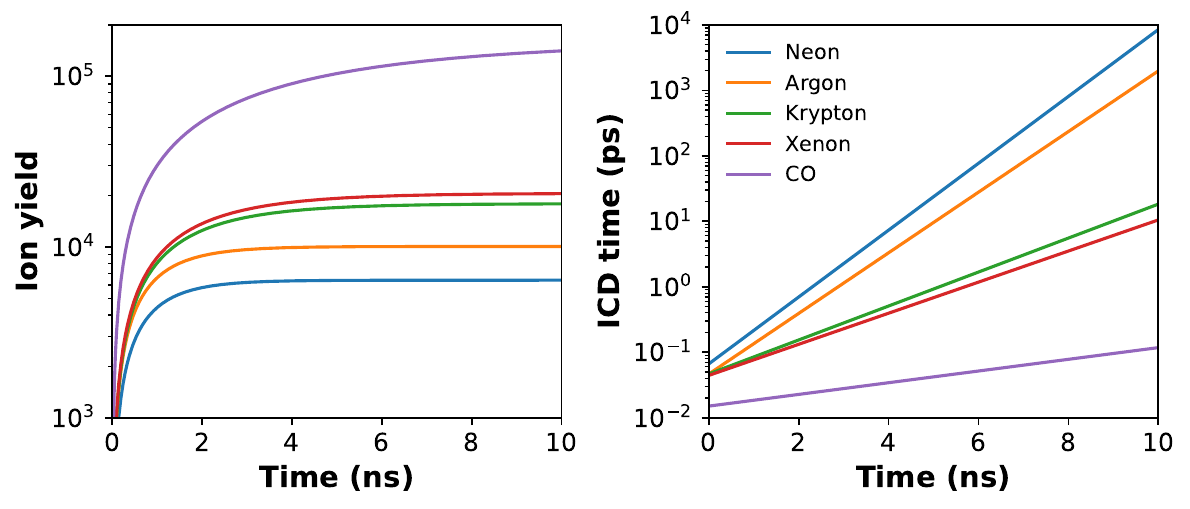}
	\caption{Ion yield (left panel) and ICD time (right panel) for Ne, Ar, Kr, Xe and CO gases as a function of time. Each gas is in a volume of 0.125~cm$^3$ and initially contains $5 \times 10^{11}$ excited species which constitute 10\% of its species. The ion yield of the Ne gas is seen to be the smallest and that of the molecular gas by about an order of magnitude larger than that of all the atomic gases. At short times, the ICD times of all the atomic gases are similar, while that of the CO gas is much shorter and stays so over time. For comparison, the radiative lifetimes are: 1.7~ns (Ne), 1.88~ns (Ar), 3.36~ns (Kr), 3.66~ns (Xe) and 9.71~ns (CO), see SI.}
	\label{Ion_Yield_All_vs_Time}
\end{figure}

We now turn to our other examples. The excited states and their data employed in the calculations are collected in Tables~S1 and S2 of the SI. The results are depicted in Figure~\ref{Ion_Yield_All_vs_Time} where the evolution of the ion yields (left panel) and ICD times (right panel) are shown for the same initial number of excited species in the gas.  It is eye catching that the molecular gas behaves differently from the atomic ones. The molecular (CO) ion yield rises much faster and saturates much later than the atomic ones. The CO yield reaches more than $10^5$ ions which is about one order of magnitude larger than that of the atoms. The four atoms can be grouped into two pairs, Ne and Ar have the smallest ion yield and Kr and Xe have a larger and very similar yield. This becomes even more evident in their ICD times. At very short times all four atoms exhibit similar values which grow in time differently for Ne/Ar and Kr/Xe. The ICD time of CO is much shorter and stays below 0.1~ps in the relevant time interval where ICD is active. The radiative lifetimes of the systems again underline the grouping: 1.7 and 1.88~ns for Ne and Ar, 3.36 and 3.66~ns for Kr and Xe, and 9.71~ns CO (see SI for details and references). The short ICD time and long radiative lifetime for CO explain its high ion yield: there is much time for ICD to be active. 

\subsection{Exciting the gas by a laser pulse}
\label{sec:Exciting by a laser pulse}

Above, we have proven that ICD is active in gases and the driving force is provided by retardation. Although these findings constitute the main result of this work, we would like to augment them by the most direct way to populate a manifold of excited species, namely by a laser pulse. We use a Gaussian pulse with peak intensity I$_0$ and a central frequency $\omega$, where $\hbar\omega$ is the excitation energy of the species in question. The modified rate equations incorporating the laser pulse are given in the Methods section, Eq.~\ref{ICD_rate_equations_laser}). In contrast to the basic scenario discussed above, all the species of the gas are initially in the ground state and the laser pulse may excite them. Since the pulse has a finite duration, this excitation is time dependent and while the excitation takes place, there is an interplay with radiative decay and ICD, leading to a rather complex overall dynamics. In addition to the excitation, the pulse may also ionize the gas by two-photon processes which is incorporated into the rate equations. The values of the respective two-photon ionzation cross sections $\sigma^{(2)}$ are rarely known but, fortunately, they are known for Ne and Ar~\cite{mckenna2003multiphoton}, see Table~S2 in SI. 

\begin{figure}[!h] 
	\centering
	\includegraphics[width=1.0\textwidth]{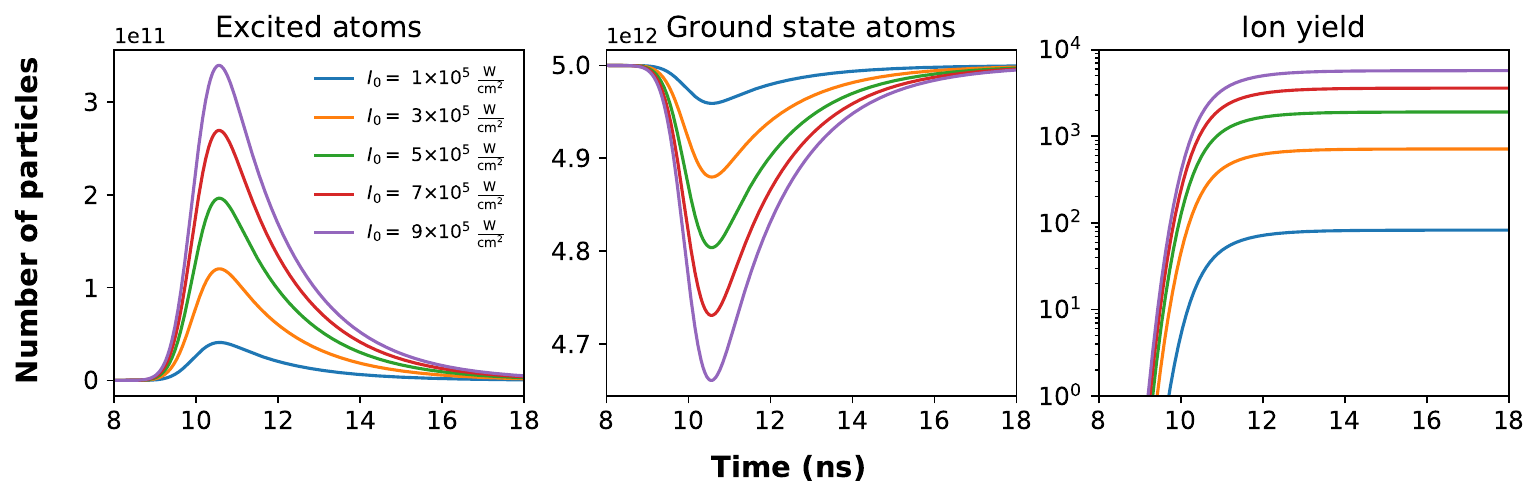}
	\caption{Time evolution of the number of excited (left panel) and ground state (middle panel) atoms as well as of the ion yield (right panel) for different peak intensities ranging from $1\times 10^{5}$ to $9\times 10^{5}$~W/cm$^2$. A Ne gas of $5\times 10^{12}$ Ne atoms in the interaction volume of 0.125~cm$^3$ (0.162~Pa pressure at room temperature) is illuminated by a Gaussian pulse of 1~ns duration. To avoid negative times, the peak of the pulse arrives at the gas after 10~ns. The radiative lifetime of an excited Ne atom is 1.7~ns. The ion yield is essentially only due to ICD.}
	\label{Excitation_Pulse_vs_Time}
\end{figure}

We have performed several calculations varying the number of ground state species (i.e., the pressure) in the interaction volume, the duration and intensity of the pulse. Figure~\ref{Excitation_Pulse_vs_Time} shows a characteristic behavior of the dynamics of the process. For a gas of $5\times 10^{12}$ Ne atoms and a pulse duration of 1~ns (the radiative lifetime is 1.7~ns), the figure depicts the evolution of the number of excited and ground state atoms as well as of the ion yield for different peak intensities between $1\times 10^{5}$ and $9\times 10^{5}$~W/cm$^2$. The center of the pulse arrives at the gas at 10~ns. As the pulse duration is shorter than the radiative lifetime, the production of excited atoms leads to asymmetric curves. Two-photon ionization is found to be rather minor and, hence, the ion yield is essentially by ICD. Since the ICD time depends strongly on the number of excited species, it is rather long (several ns) at the rise of the pulse, short (0.1 to 10~ps) at the peak and growing long again exponentially, see Figure~S1 in SI. This behavior explains the less steep rise of the ion yield than found in Figure~\ref{Number_Particles_vs_Time}.

\begin{figure}[!h] 
	\centering
	\includegraphics[width=1.0\textwidth]{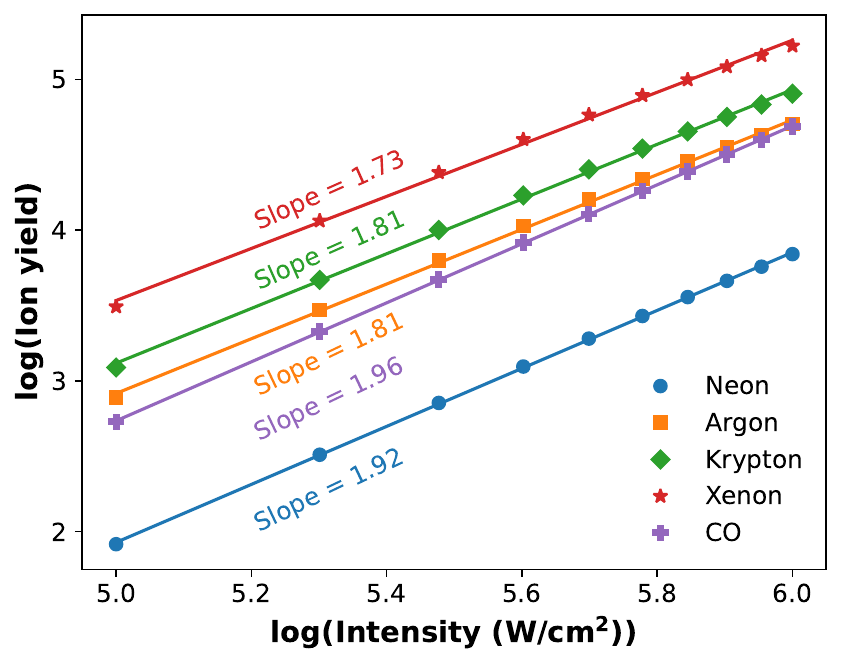}
	\caption{A log-log plot of the ion yield versus the peak intensity of the Gaussian pulse of 1~ns duration for gases of Ne, Ar, Kr, Xe and CO. Each of the gases contains $5\times 10^{12}$ species in the interaction volume of 0.125~cm$^3$.}
	\label{Log_Yield_vs_Log_Intensity_All}
\end{figure}

  As our last result we show in Figure~\ref{Log_Yield_vs_Log_Intensity_All} a log-log plot of the ion yield as a function of peak intensity for all studied systems. As the  two-photon ionization cross sections $\sigma^{(2)}$ are only known for Ne and Ar, we took the value for Ne also for Kr, Xe and CO. As mentioned above, the impact of this quantity is minor for Ne and Ar. The yield is by far the smallest for the Ne gas and largest for Xe. Clearly, the ion yield increases with increasing peak intensity. As the intensity studied is rather low, the slopes of the log-log plots are close to 2 because the ICD requires two excited species for each event and two photons are also involved in two-photon ionization. As the intensity grows, the curves start to bend due to saturation effects and this reduces the slope. The bending can already be seen by the eye for Xe. 

\section{Conclusion} \label{sec:Conclusion}

There is ample theoretical and experimental investigations on ICD in weakly bound systems like in clusters and fluids. In the present work, we pose the general question whether ICD can also be active in gases, and if yes, what is the underlying mechanism. In weakly bound systems, the interspecies distances typically amount to several $\text{\AA}$ up to a few nm. In gases, the scale is rather $\mu$m. On this length scale, we show that the Coulomb interaction which drives ICD in weakly bound systems is negligible. We demonstrate here that retardation effects which arise from the finite speed of light constitute the mechanism to make ICD operative in gases and that the production of ions by ICD can be rather impressive. At first sight, this may be surprising as retardation has been found to lead to very small effects in weakly bound systems, and in gases where the distances are much larger one might expect them to be even less relevant. ICD has many modes of appearance. In ICD of multiply excited species, an excited atom or molecule transfers its excess energy to another excited species and ionizes it. Each of these species can play the role of a donor as well as of an acceptor. The impact of retardation grows very favorably when increasing the number of excited species even if the volume containing them increases too. 

We show that there is a competitive interplay between the radiative decay of the excited species and the decay by ICD. As the ICD rate increases if $\gamma_{rad}$ is increased (see Eqs.~(\ref{ICD_rate_Coulomb}) and (\ref{ICD_rate_Retardation})), choosing systems or states with larger radiative rate by no means implies less ICD.  The radiative rate stays constant in time, but as the number of excited species varies in time, the impact of ICD does change in time. This all leads to rather intricate dynamics of the production of ions by ICD. If the excited species are produced by a laser pulse, the dynamics becomes even more involved as there is, in addition to the above, competition between excitation and decay~\cite{Competition_Excitation_Decay}. By varying the pulse parameters (intensity, duration, frequency), one can manipulate the dynamics and control the ion production by ICD. 

Having proven that ICD is operative in gases and understood its basic mechanism considerably broadens the scope of ICD and opens up new dimensions to applications and understanding of phenomena occurring in nature, like in planet atmospheres, in dense interstellar clouds and growth of molecules in space~\cite{ICD_in_unbound_pyridines,ICD_in_unbound_PANH}. The present results also shed light on other cases where retardation may be decisive too, like collective ICD and energy transfer in general.  

 
\section*{Methods} \label{sec:Methods}

\subsection*{Basic scenario}
Consider an isotropic rarefied gas of volume $V$ composed of atoms or molecules, $N$ of them being in their ground state and $N^{*}$ in an electronically excited state, as illustrated in Figure~\ref{Schematic_Cartoon_ICD_Gas}. The excited species may decay either radiatively $A^* \rightarrow A + \hbar \omega$ by emitting a photon at the rate $\gamma_{rad}$ or by ICD via the pairwise mechanism of multiply excited species~\cite{PhysRevLett.105.043004} $A^* + A^* \rightarrow A + A^+ + e_{_{ICD}}$ emitting an electron and resulting in the production of $N^{+}$ ionized species. All these particle numbers evolve in time.
 
The ICD rate for an energy transfer from an excited species $j$ to an excited species $i$ separated by a distance $R_{ij}$ is given by (in atomic units)~\cite{Retardation_ICD_1,Relativistic_Energy_Transfer}
\begin{equation}\label{ICD_rate_pair}
	\gamma_{_{ICD}} (i,j) =  \frac{\gamma_{rad} \sigma_{pi}^{*}}{4 \pi} \left[3 \left(\frac{c}{\omega}\right)^4 \frac{1}{R_{ij}^{6}} + \left(\frac{c}{\omega}\right)^2 \frac{1}{R_{ij}^{4}} + \frac{1}{R_{ij}^{2}}\right] e^{-\frac{R_{ij}}{R_{att}}},
\end{equation}
where $\sigma_{pi}^{*}$ is the photoionization cross section of the excited state, $c$ is the speed of light, and $\omega$ is the excess energy transferred via a virtual photon. Similarly to the situation when a beam of photons passes through matter and its intensity decreases exponentially with distance owing to
absorption and scattering, we must also include attenuation by the medium, as particles in the gas can absorb the virtual photons~\cite{Retardation_ICD_1,Relativistic_Energy_Transfer}. In our case, the gas constituents are the species in the ground and excited states as well as the produced cations. Under the present conditions of low density we may neglect the impact of the cations produced and, consequently, the attenuation length $R_{ att}$ takes on the following expression
\begin{equation}\label{Attenuation}
	\frac{1}{R_{att}} \approx \rho_{g} \sigma_{abs} + \rho^* \sigma_{pi}^*
\end{equation}
where $\sigma_{abs}$ is the photoabsorption cross section of the species in their ground state at the transferred energy $\omega$, and $\rho_{g}$ and $\rho^{*}$ are the densities of the species in the ground and excited states, respectively. The first term on the r.h.s. of the above equation describes the absorption of photons of energy $\omega$ by the species in the ground state, and being resonant with the excitation energy of these species, this term dominates by far the contribution to the attenuation length. This statement becomes even more obvious considering that the density $\rho_{g}$ is at least 10 times larger than $\rho^{*}$ in our examples. We note that the attenuation length varies in time as the densities change in time due to radiative emission and ICD, i.e., $\rho_{g}$ increases and $\rho^{*}$ decreases as time proceeds.

As we have a large number of particles, it is hardly possible to compute the time evolution of the system taking into account all the particles individually and it is convenient to introduce rate equations. To that end we bring in the total ICD rate $\Gamma_{ICD}$ of the setup which is a sum of the ICD rates  of all pairs of excited particles. The total ICD rate reads
\begin{equation}\label{Total_ICD_rate}
	\Gamma_{ICD} = \frac{1}{2} \sum_{i = 1}^{N^{*}} \sum_{\substack{j = 1 \\ j \neq i}}^{N^{*}} \gamma_{_{ICD}} (i,j),
	\end{equation}
where the factor $1 / 2$ avoids double counting, as each pair of excited particles contributes twice in the sum. As the number $N^{*}$ of excited species varies in time, the total rate is a temporal quantity describing the evolution of ICD in the whole gas.

Although the above expression for the total ICD rate is simple, its numerical evaluation is very involved as this quantity scales as $(N^{*})^2$ and varies in time, i.e., has to be evaluated many times. In our examples $N^{*}\approx10^{11}$. As already mentioned in the former section,  to enable the calculation we have distributed the excited species isotropically in the volume, for instance, by choosing a cube of volume V and placing the species on the sites of a simple cubic lattice. Then, the shortest distance between two excited species is the average distance of a gas of excited species. This distance grows further in due time as the number of excited species decreases. Therefore, there are no fluctuations in the gas which can bring excited species close to each other, and we may consider our computed ion yield to be a lower bound.  As a consequence, one can now approximate the summation in Eq.~\eqref{Total_ICD_rate} by an integration over the coordinates of a pair of excited species
\begin{equation}\label{Total_ICD_rate_integral}
	\Gamma_{ICD}  \approx  \frac{(\rho^{*})^2}{2} \iiint d^3 r_1 \iiint d^3 r_2 \gamma_{ICD} (1,2),
\end{equation}
where the density of excited species $\rho^{*} = N^{*} / V$ is a constant in the spatial coordinates space, but varies in time since $N^{*} = N^{*} (t)$. The integral can be solved analytically by considering a sphere of the same volume $V$ as the cube (with sides $2L$). 
By using Eq.~\eqref{ICD_rate_pair}, the total ICD rate can be expressed explicitly as
\begin{equation}\label{Total_ICD_rate_final}
	\Gamma_{ICD}(N^{*}) \approx \frac{\gamma_{rad} \sigma_{pi}^{*}}{8 \pi V^2} \Biggl[3\left(\frac{c}{\omega}\right)^{4} I_6
	+ \left(\frac{c}{\omega}\right)^{2} I_4
	+ I_2 \Biggr] (N^{*})^2
\end{equation}
where
\begin{equation}\label{The_3_integrals}
	\begin{split}
		I_n  = \left(\frac{L}{\alpha}\right)^{6 - n}
		\Biggl\{\Biggl\{& \Tilde{\Gamma}\Biggl[6 - n, \alpha \left(\frac{6}{\pi}\right)^{1 / 3}\Biggr] - \Tilde{\Gamma}\Biggl[6 - n, \alpha \frac{\langle R \rangle}{L}\Biggr] \Biggr\}\\
		- 16 \alpha \Biggl\{ & \Tilde{\Gamma}\Biggl[5 - n, \alpha \left(\frac{6}{\pi}\right)^{1 / 3}\Biggr] - \Tilde{\Gamma}\Biggl[5 - n, \alpha \frac{\langle R \rangle}{L}\Biggr]\Biggr\}\\
		+ 24 \alpha^2 \pi \Biggl\{ & \Tilde{\Gamma}\Biggl[4 - n, \alpha \left(\frac{6}{\pi}\right)^{1 / 3}\Biggr] - \Tilde{\Gamma}\Biggl[4 - n, \alpha \frac{\langle R \rangle}{L}\Biggr] \Biggr\}\\
		- 32 \alpha^3 \pi  \Biggl\{ & \Tilde{\Gamma}\Biggl[3 - n, \alpha \left(\frac{6}{\pi}\right)^{1 / 3}\Biggr] - \Tilde{\Gamma}\Biggl[3 - n, \alpha \frac{\langle R \rangle}{L}\Biggr]\Biggr\} \Biggr\}.
	\end{split}
\end{equation}
where $\Tilde{\Gamma}(m, x)$ is the incomplete Gamma function~\cite{abramowitz1965handbook}, $\alpha = L / R_{att}$, and $\langle R \rangle$ is the nearest neighbor distance of excited particles in the gas, given by~\cite{chandrasekhar1943stochastic}
\begin{equation}\label{Nearest_neighbor_distance}
	\langle R \rangle = \left(\frac{3}{4 \pi \rho^{*}}\right)^{1/3} \Gamma \left(\frac{4}{3}\right),
\end{equation}
where $\Gamma \left(4/3\right) \approx 0.8929795$.

We are now in the position to model the time evolution of the gas starting with given initial numbers $N(0)$ of species in the ground state, ${N}^{*}(0)$ of excited ones and ${N}^{+}(0)$ of ions. The corresponding set of coupled rate equations takes on the following appearance: 
\begin{equation}\label{ICD_rate_equations}
	\begin{aligned}
		\frac{dN^{*}}{dt} & = \dot{N}^{*}  = - \gamma_{rad} N^{*} - 2 \Gamma_{ICD}(N^{*}), \\
		\frac{dN}{dt} & = \dot{N} = + \gamma_{rad} N^{*} +   \Gamma_{ICD}(N^{*}), \\
		\frac{dN^{+}}{dt} & = \dot{N}^{+} =                         + \Gamma_{ICD}(N^{*}). 
	\end{aligned}
\end{equation}
All the above particle numbers $N(t)$, $N^{*}(t)$, and $N^{+}(t)$ vary in time as is also the case for $\Gamma_{ICD}(N^{*})$, and as the total number of species $N_{tot}$ is conserved, $N_{tot} = N(t) + N^{*}(t) + N^{+}(t) = \text{constant}$. Note that $\Gamma_{ICD}(N^{*})$ changes in time not only due to its quadratic dependence on $N^{*}$, but also owing to the fact that $R_{att}$ and the nearest neighbor distance of excited particles in the gas $\langle R \rangle$ which enter the expression for this rate change in time, see Eqs.~\eqref{Total_ICD_rate_final}, \eqref{The_3_integrals}, \eqref{Nearest_neighbor_distance} and \eqref{Attenuation}.

In the calculations presented in subsection `The basic scenario', the initial number of ions is zero and all ions found at later times are due to ICD. The excited species can decay by radiative emission producing a ground state one $A^*\rightarrow A$ and a photon (not explicitly shown) and also by ICD $A^* + A^* \rightarrow A + A^+$ also producing a ground state one and an ion accompanied by an emitted electron (not explicitly shown). These two decay pathways are indicated by the terms $- \gamma_{rad} N^{*}$ and $- 2 \Gamma_{ICD}(N^{*})$ on the r.h.s. of the first rate equation of Eq.~\eqref{ICD_rate_equations}, respectively. The factor 2 stems from the fact that two excited species are lost in every ICD event. Owing to the two pathways, the number of ground state species grows with every radiative decay and ICD event and that of the ionized species only by an ICD event. Finally, we remind that $\Gamma_{ICD}(N^{*})$ is proportional to $(N^*)^2$, see Eq.~\eqref{Total_ICD_rate_final}, implying that ICD can be very efficient if many excited species are present.

The set of differential equations~\eqref{ICD_rate_equations} has been solved numerically employing the Runge-Kutta 4th order method~\cite{butcher2016numerical}. Thereby, the initial conditions at $t  = 0$ chosen were: $N^{*} (0) = 0.1 N_{tot}$, $N(0) = 0.9 N_{tot}$ and $N^{+}(0) = 0$, i.e, 10\% of the species were excited and 90\% in the ground state. The total number $N_{tot}$ in the volume $V = 0.125$~cm$^3$, which matches the interaction volume reported in the experiment of Ref.~\cite{ICD_in_unbound_pyridines}, has been varied from $1 \times 10^{12}$ to $9 \times 10^{12}$. The ICD rate $\Gamma_{ICD}$ was computed at each time step using Eqs.~(\ref{Total_ICD_rate_final}) and (\ref{The_3_integrals}). The values of the input parameters $\omega$, $\gamma_{rad}$ and $\sigma_{pi}^{*}$ are collected in Table~S2 of the SI. 

\subsection*{Inclusion of the laser pulse}

As mentioned above, the main goal of this work is to prove that ICD takes place in a gas in spite of the large distances between the species. Nevertheless, we would like to draw a connection to a possible experiment where the excited species are produced by a laser pulse. To that end we consider a gas where all particles are initially in the ground state which interacts with a laser pulse of intensity $I(t)$ and assume that the laser pulse overlaps with the gas in an interaction volume $V$. The central frequency of the pulse is selected to be resonant with the excitation energy of the electronic level of interest (see Table~S1 of the SI). In the following we consider only the gas particles in the interaction volume and put $N_{tot}=N(0)$. Clearly, the pulse excites ground state species only during its duration, but in addition, it also deexcites already excited species by stimulated emission and this interplay has to be taken into account in Eq.~(\ref{ICD_rate_equations}) following the well-known rate equations of Ref.~\cite{milonni2010laser}. Furthermore, we have to include the possibility of resonant two-photon ionization~\cite{lambropoulos1974theory,mckenna2003multiphoton}. The resulting augmented rate equations now read (in atomic units)
\begin{equation}\label{ICD_rate_equations_laser}
	\begin{aligned}
		\dot{N}^{*} & = - \gamma_{rad} N^{*}  - 2 \Gamma_{ICD}(N^{*}) + \frac{\sigma_{abs}}{\omega} I (t) \left(N - N^{*}\right),  \\
		\dot{N} & =   + \gamma_{rad} N^{*}  + \Gamma_{ICD}(N^{*}) - \frac{\sigma_{abs}}{\omega} I(t) \left(N - N^{*}\right) - \sigma^{(2)} \left[\frac{I(t)}{\omega}\right]^2 N,   \\
		\dot{N}^{+} & =  +  \Gamma_{ICD}(N^{*}) + \sigma^{(2)} \left[\frac{I(t)}{\omega}\right]^2 N  .
	\end{aligned}
\end{equation}
The term $(\sigma_{abs}/\omega) I(t) \left(N - N^{*}\right)$ accounts for absorption and stimulated emission in the system~\cite{milonni2010laser}. As we consider pulses which excite up to about 10\% of the species, this term essentially continuously increases the number of excited and reduces the number of ground state species. The term $\sigma^{(2)} \left[I(t)/\omega\right]^2 N$, where $\sigma^{(2)}$ is the two-photon ionzation cross section~\cite{mckenna2003multiphoton}, describes resonant two-photon ionization which for a monochromatic and coherent laser pulse like used here is a one-step process~\cite{lambropoulos1974theory}.

The set of differential equations~\eqref{ICD_rate_equations_laser} has been solved numerically employing the Runge-Kutta 4th order method~\cite{butcher2016numerical}. The initial conditions chosen were $N(0) = N_{tot}$ and $N^{*} (0) = N^{+}(0) = 0$, i.e., all species were in the ground state, and their total $N_{tot} = 5 \times 10^{12}$ in the laser-gas interaction volume ($V = 0.125$~cm$^3$) was kept fixed similar to that in~\cite{ICD_in_unbound_pyridines}. 

The laser pulse intensity $I(t)$ has been modeled as a full-width-at-half-maximum Gaussian pulse
\begin{equation}\label{Pulse_Gaussian}
	I (t) = I_0 \exp \left[- 4 \ln{2} \frac{(t - t_0)^2}{t_p^2}\right],
\end{equation}
where $I_0$ is the peak intensity of the pulse, $t_0$ the time at which the pulse is applied, and $t_p$ is the duration of the pulse. We selected typical values for the laser parameters based on Ref.~\cite{ICD_in_unbound_pyridines} and varied $I_0$ from $1 \times 10^5$ to $9 \times 10^5$~W/cm$^2$, maintaining the pulse width $t_p$ of $1$~ns. For convenience, the pulse center was settled to $t_0 = 10$~ns. The values of the input parameters $\omega$, $\gamma_{rad}$, $\sigma_{abs}$, $\sigma_{ pi}^{*}$ and $\sigma^{(2)}$ are collected in Table~S2 of the SI.

\section*{Acknowledgments}
The authors thank S. Buhmann for fruitful discussions. Financial support by the Deutsche Forschungsgemeinschaft (DFG) (Grant No. CE 10/56-1) is gratefully acknowledged. A.G.F. acknowledges support from grants Nrs. 2024/17762-5 and 2025/07261-1, São Paulo Research Foundation (FAPESP). This research used the computing resources and assistance of the John David Rogers Computing Center (CCJDR) in the Institute of Physics “Gleb Wataghin”, University of Campinas.

\section*{Data Availability}
The data that supports the findings of this study are available from the corresponding author upon reasonable request.

\bibliography{references_Relativistic_Dark.bib}
\bibliographystyle{naturemag}

\end{document}